\documentclass[twoside,12pt,a4paper]{article}
\usepackage[dvips]{epsfig}
\usepackage{a4}
\usepackage{latexsym}
\usepackage{graphics}
\def\beq{\begin{equation}}
\def\eeq{\end{equation}}
\def\beqn{\begin{eqnarray}}
\def\eeqn{\end{eqnarray}}

\begin{document}

\title{Helicity Amplitudes \\ 
In the Hypercentral Constituent Quark Model}

\author{M.M. Giannini, E. Santopinto, A. Vassallo\\
\small{
Dipartimento di Fisica dell'Universit\`a di Genova, I.N.F.N. 
Sezione di Genova, Italy}}

\maketitle

\abstract{
We report on the recent results of the hypercentral constituent
quark model\cite{pl,iso}. The model contains a spin independent three-quark
interaction which is inspired by QCD lattice calculations and reproduces
the average energy values of the $SU(6)$ multiplets.  The splittings are
obtained with a residual generalized $SU(6)$ -breaking interaction,
including an isospin dependent term \cite{iso}.
The long standing problem of the Roper resonance is absent and all
the 3- and 4-star states are well reproduced. 
The model has also been used for predictions concerning the electromagnetic
transition form factors giving a good description of the medium 
$Q^2$ -behaviour \cite{aie,aie2}. In particular the calculated 
$S_{11}~~A_{\frac{1}{2}}$ helicity amplitude agrees very well with the recent
CLAS data \cite{burk}. 
Finally the ratio of the elastic form factors of
the proton \cite{rap}, calculated including kinematic relativistic
corrections, exhibits a substantial decreasing with $Q^2$
in agreement with the recent TJNAF experiment \cite{ped} .}

\section{The Hypercentral Model}
\label{subsec:prod}
The model \cite{pl} consists of a hypercentral quark interaction containing a 
linear plus coulomb-like term as suggested 
by lattice QCD calculations\cite{bali}. It can be considered as the
hypercentral 
approximation of the two-body potential or as a three-body potential
\begin{equation} 
V(x)= -\frac{\tau}{x}~+~\alpha x~~~~, ~~~\mathrm{with} ~~
~x=\sqrt{\mbox{\boldmath{$\rho$}}^2+{\mbox{\boldmath{$\lambda$}}}^2} ~~, 
\end{equation} 
\noindent 
where $x$ is the hyperradius defined in terms of the Jacobi
coordinates 
$\mbox{\boldmath{$\rho$}}$ and $\mbox{\boldmath{$\lambda$}}$ .
A hyperfine term of the standard
form \cite{is} is added and treated as a perturbation.
After having fixed the quark mass $m$ to $1/3$ of the nucleon mass, the
average energies of the $SU(6)$-multiplets are described  with $\tau=4.5$
and $\alpha=1.61~fm^{-2}$, while the strength of the
hyperfine
interaction is determined by the $\Delta$ - Nucleon mass difference. The
wave functions of the various resonances are therefore
completely determined (the few parameter of the model fixed once for all
at the reproduction of the spectrum) and they have been used for the
calculation
of the photocouplings \cite{aie}, the transition form factors to the
negative parity resonances \cite{aie2}, the elastic form factors
\cite{mds} and the ratio between the electric and magnetic form factors of
the proton \cite{rap}.
\section{Electromagnetic transition end elastic form factors}

The baryon spectrum is usually described quite well by various Constituent
Quark Models \cite{pl,is,ci,olof}, although the various models are 
quite different. In order to distinguish among the various forms of quark 
dynamics one has to study in a consistent way all the physical observables
of interest and not only the spectrum which is a static property.
The helicity amplitudes for the electroexcitation of baryon resonances,
are calculated using the states determined by the model and a non
relativistic current for point quarks.

\begin{figure}[ht]
\hspace*{-2truecm}
\begin{center}
\includegraphics[scale=1.0] {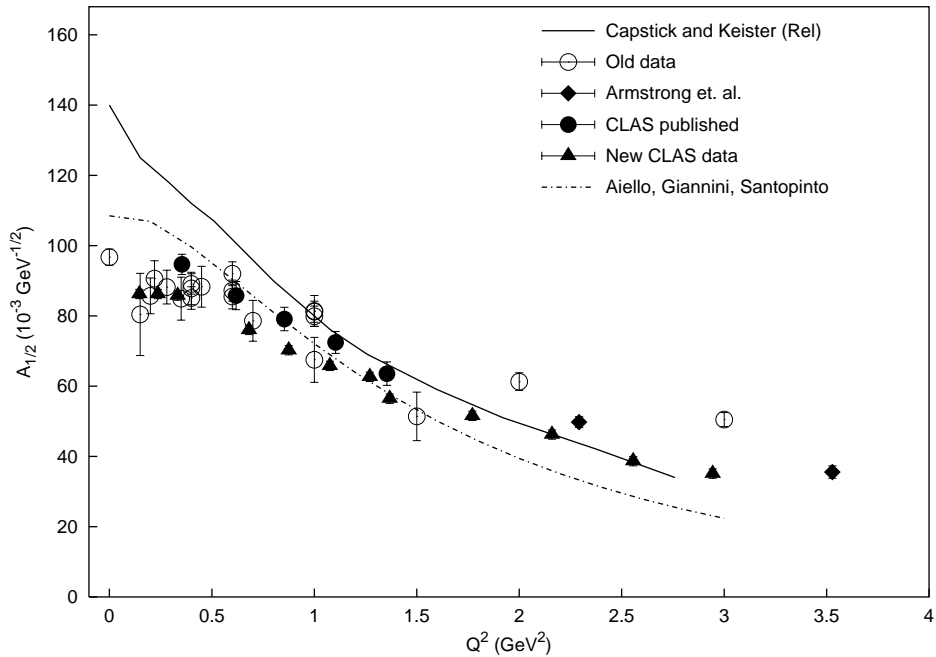}
\caption{\footnotesize 
Comparison between the experimental data \cite {burk} for the 
helicity amplitude $A^p_{1/2}$ for the $S_{11}(1535)$
resonance and the calculations with the hCQM, lower curve \cite{aie2} 
also compared with Capstick and Keister result, upper curve \cite{capstick}.}
\end{center}
\end{figure}

\noindent In Fig. 1 we report the helicity amplitude for the $S_{11}(1535)$
resonance. Similar results are obtained for the remaining
negative parity nucleon resonances \cite{aie2} and also in a systematic way
for all the other 3-4 star and 1-2 star resonances \cite{aie3}.

In general the $Q^2$ behaviour is reproduced, except for
discrepancies at small $Q^2$, especially in the
$A^{p}_{3/2}$ amplitude of the transition to the $D_{13}(1520)$ state. 
These discrepancies could be ascribed to the non-relativistic character of
the model, and to the lack of explicit quark-antiquark configurations 
which may be important at low $Q^{2}$. 
The kinematical relativistic corrections at the level of boosting the nucleon 
and the resonances states to a common frame are not 
responsible for these discrepances,  as we have demonstrate in 
 Ref.\cite{mds2}.

These boosts effects are on the contrary important for the elastic e.m. form
factors. Taking into account the boosts of the
 3-quark states for the nucleon from the rest frame to the Breit frame one
can write   
\begin{equation}\label{eq:gmff} 
G_{E}(Q^2) = F^C_{el} G_{E}^{nr} (q/g)~, 
~~~G_{M}(Q^2) =  F^M_{el} G_{M}^{nr} (q/g)~, 
\end{equation} 
where $G_{E}^{nr}$, and $G_{M}^{nr}$ are the electric and magnetic form
factors as given by the non relativistic quark model, $F^C_{el}$ and
$F^M_{el}$ are kinematical factors and $g~=~E/M$.  The formula of Eq.2 can
be used for any CQMs \cite{mds,mds2}.In particular,
the elastic form factors of Eq. (\ref{eq:gmff}), calculated using as input
the nucleon form factors obtained in the hCQM, lead to    
an improvement of the theoretical description \cite{mds,rap}, especially
the shape of the elastic form factors as a function of $Q^2$
is similar to that of the experimental data. 
In Fig. 2 we report the ratio
of the electric and magnetic
form factors of the
proton, $R~=~\mu_p~\frac{G_{E}(Q^2)}{G_{M}(Q^2)}$ \cite{rap}.

\begin{figure}[ht]
\begin{center}
\includegraphics[scale=.65,angle=90]{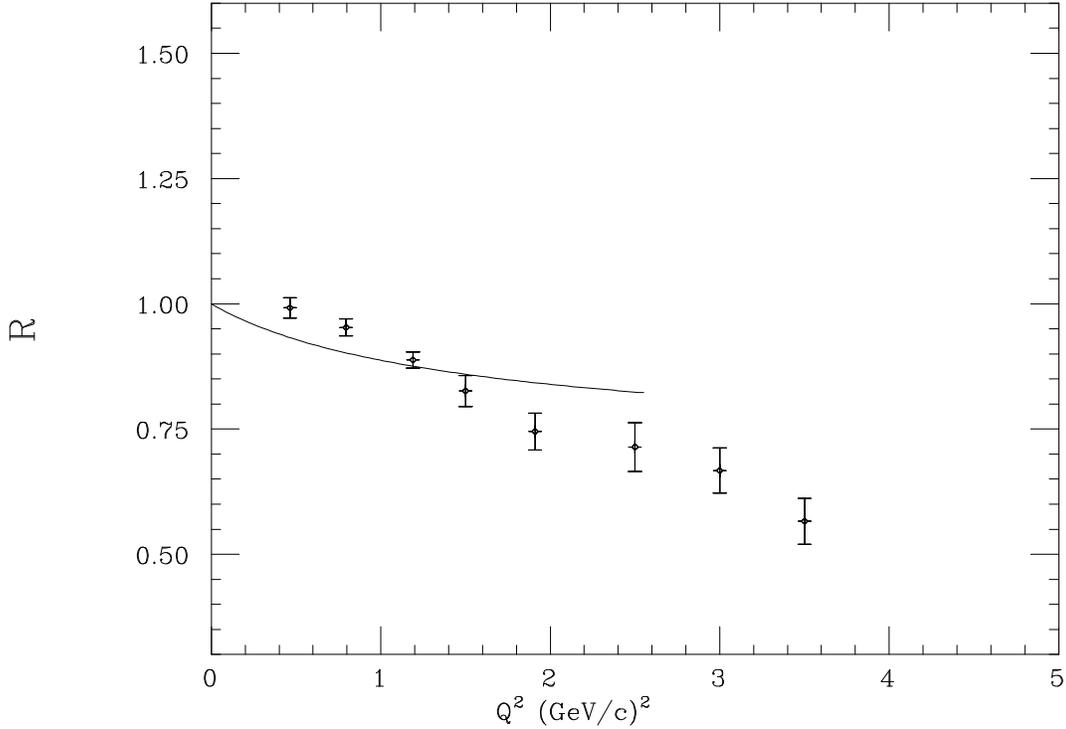}
\caption{\footnotesize The ratio $R~= {\mu_p}~G_{E}/G_{M}$  calculated with
the hCQM \cite{rap}. The points are the data from a recent TJNAF experiment 
\cite{ped} } 
\end{center}
\end{figure} 

The data of a recent polarization transfer experiment at TJNAF show a    
significant deviation from the 
scaling behaviour, which is reproduced up to $1.5~(GeV/c)^2$ by the hCQM  
model calculation \cite{rap} (full curve in Fig. 2).  It should be reminded
that the non relativistic calculation
gives $R=1$ and it remains 
 $1$ within $1\%$ even if the hyperfine mixing
is included.  The decreasing of the ratio $R$ with $Q^2$  
is due to the different behaviour of the relativistic correction for the 
electric and magnetic parts. 

\section{Generalized SU(6)- breaking interaction}

There are different motivations for the
introduction of a flavour dependent term in the three-quark interaction.
The well known Guersey-Radicati mass formula \cite{gura}
contains a flavour dependent term, which is essential for the description
of the strange baryon spectrum.
In the chiral Constituent Quark Model \cite{olof,ple}, the non
confining part of the   
potential is provided by the interaction with the Goldstone bosons,
giving rise to a spin- and isospin-dependent part, which is crucial in
this approach for the description of the lower part of the spectrum.
More generally, one can expect that the quark-antiquark pair production 
can lead to an effective residual quark interaction containing an isospin
(flavour) dependent term.
We have introduced isospin dependent terms
in the hCQM hamiltonian.
The complete interaction used is given by\cite{iso}
\begin{equation}\label{tot}
H_{int}~=~V(x) +H_{\mathrm{S}} +H_{\mathrm{I}} +H_{\mathrm{SI}}~,
\end{equation}
where $V(x)$ is the linear plus hypercoulomb SU(6)-invariant potential,
 while $H_{\mathrm{S}} + 
H_{\mathrm{I}} +H_{\mathrm{SI}}$ is a residual SU(6)-breaking interaction,
with ${H}_{\mathrm{S}}$ a smeared standard hyperfine term,
${H}_{\mathrm{S}}$ a spin dependent term,
${H}_{\mathrm{I}}$ isospin dependent and  ${H}_{\mathrm{SI}}$ 
spin-isospin dependent.

\begin{figure}
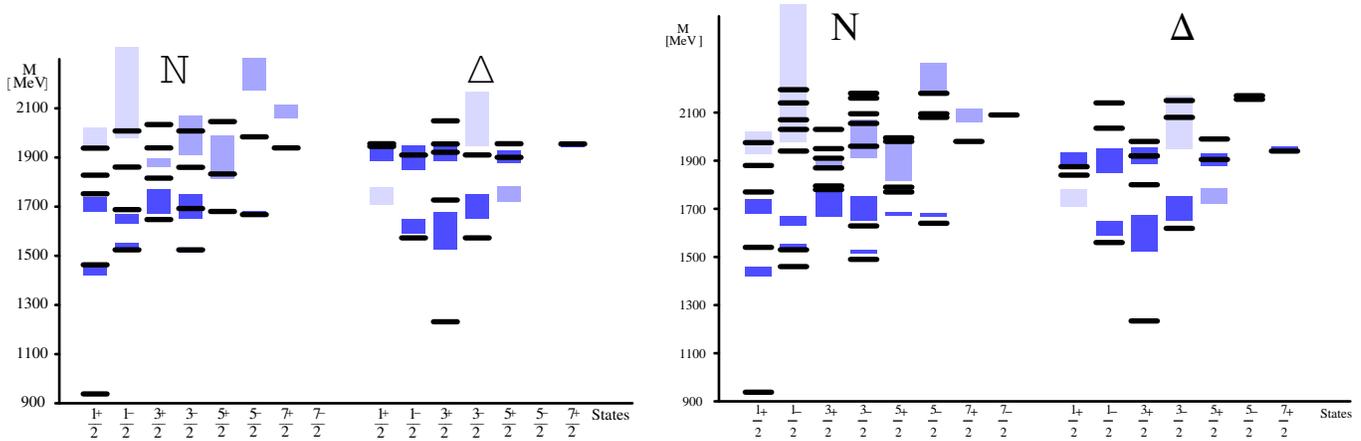

\hspace*{-2truecm}
\begin{tabular}{cc}
\includegraphics[scale=.6]{Missing_hCQM.epsi}& 
\includegraphics[scale=.5]{capstick.fig.epsi}
\end{tabular}
\caption{\footnotesize On the left the non strange spectrum obtained with the hCQM (complete
interaction (\ref{tot})). The shadowed boxes represent
the experimental
data from PDG with their uncertainty \cite{pdg}: the dark grey boxes for the 3- and 4-star
resonances
 and the light grey boxes for the 1- and 2-stars. 
On the right are reported the results of the Isgur-Capstick 
model}
\end{figure}
\noindent 
The resulting spectrum for the 3*- and 4*- resonances is shown in Fig.3
\cite{iso}.
The $N-\Delta$ mass difference is no more due only to the
hyperfine interaction, which contribute about $35\%$,
while the remaining splitting comes from the
spin-isospin term, $(50\%)$, and from the isospin one, $(15\%)$.

\section{Conclusions}
We have presented various results predicted by
 the hypercentral Constituent Quark Model  compared with the
experimental data.
We have also shown that in the hCQM a flavour dependent
potential can be introduced leading
to improved splittings within the $SU(6)$-multiplets.
A relativistic description of the dynamical properties of the nucleon 
is important and inevitable in particular for the electromagnetic form
factors.


\begin{thebibliography}{0}

\bibitem{pl}
M. Ferraris, M.M. Giannini, M. Pizzo, E. Santopinto and L. Tiator, Phys. Lett.
{\bf B364}, 231 (1995).

\bibitem{iso}
M.M. Giannini, E. Santopinto and A. Vassallo, Eur. Phys. J. {\bf A12}, 447
(2001); M.M. Giannini, E. Santopinto and A. Vassallo,
Nucl. Phys. {\bf A699}, 308(2002).

\bibitem{aie}
M. Aiello, M. Ferraris, M.M. Giannini, M. Pizzo and E. Santopinto,
Phys. Lett {\bf 387}, 215 (1996). 

\bibitem{aie2}
M. Aiello, M. M. Giannini, E. Santopinto, J. Phys. G: Nucl. Part. Phys.
{\bf 24},
753 (1998)
 
\bibitem{burk}
V.~D.~Burkert,arXiv:hep-ph/0207149. 

\bibitem{rap}
M. De Sanctis, M.M. Giannini, L. Repetto, E. Santopinto, Phys. Rev.
{\bf C62},025208 (2000).

\bibitem{ped}
M.K. Jones et al., Phys. Rev. Lett. {\bf B84},1398 (2000).

\bibitem{bali}
Gunnar S. Bali, Phys. Rep. {\bf 343}, 1 (2001).

\bibitem{is}
N. Isgur and G. Karl,
Phys. Rev. {\bf D18}, 4187 (1978); {\bf D19}, 2653 (1979); {\bf D20}, 11
(1979); S. Godfrey and N. Isgur, Phys. Rev. {\bf D32}, 189 (1985). 

\bibitem{mds}
M. De Sanctis, E. Santopinto, M.M. Giannini, Eur. Phys. J. {\bf A1}, 187 
 (1998).

\bibitem{ci}
S. Capstick and N. Isgur, Phys. Rev. {\bf D 34},2809 (1986). 

\bibitem{olof} 
L. Ya. Glozman and D.O. Riska, Phys. Rep. {\bf C268}, 263 (1996).   

\bibitem{capstick}
S. Capstick and B.D. Keister, Phys. Rev.{\bf D 51}, 3598 (1995). 


\bibitem{aie3}
M.M. Giannini, E.Santopinto, M. Aiello, to be published.


  
\bibitem{mds2} 
M. De Sanctis, E. Santopinto, M.M. Giannini, Eur. Phys. J. {\bf A2}, 
403 (1998). 
 

\bibitem{gura}
F. Guersey and L.A. Radicati, Phys. Rev. Lett.  {\bf 13}, 173 (1964).

\bibitem{ple}
L. Ya. Glozman, Z. Papp, W. Plessas, K. Varga, R. F. Wagenbrunn,
Phys. Rev. {\bf C57}, 3406 (1998).

\bibitem{pdg}
Particle Data Group, Eur. Phys. J. {\bf C15}, 1 (2000).   


\end{thebibliography}
\end{document}